# Title: Direct determination of the spin-polarization at buried interfaces using voltage dependent MOKE signals.

**Authors:** Michael Miller[1], Rene Nacke[1], Sven Ilse[1,2], Gisela Schütz[1], and Eberhard Goering[2*]

**Affiliations:**

[1]Max Planck Institute for Intelligent Systems, Modern Magnetic Systems, Heisenbergstraße 3, 70569 Stuttgart, Germany

[2]Max Planck Institute for Solid State Research, Solid State Spectroscopy, Heisenbergstraße 1, 70569 Stuttgart, Germany

*Corresponding author. Email: e.goering@fkf.mpg.de

**Abstract:** Here, we report a novel and conceptually straightforward technique to detect the spin-polarization directly, with laser spot spatial resolution, at buried ferromagnet/insulator interfaces in ambient settings. This has been accomplished by monitoring the voltage-induced change of the longitudinal MOKE signal as a function of the applied magnetic field and applying an AC voltage across the interface. For the case where the spin polarization enters the VMOKE signal, a simple quantitative model is proposed. A distinct positive majority spin polarization has been found for Fe and Co, whereas Ni exhibits a negative minority spin polarization.

**Introduction:**

For almost all spin electronic (spintronic) devices, the spin-polarization at a given interface is a key property. In the case of tunneling magneto resistance (TMR) and giant magneto resistance (GMR) sensors, the observed magnetoresistive effect is correlated to the spin-polarization at the interface as formulated in Jullier's model [1], which has been improved by Slonczewski [2]. In addition, the tunnel barrier material has revealed a strong influence on the observed TMR effect. For example, for CoFe alloys, TMR in the range of 50% has been observed using $Al_2O_3$ as a barrier [3], while MgO shows TMR up to 220% at room temperature (RT) [4]. By measuring the effective TMR, the so-called tunneling spin-polarization could be determined, for example, for Co up to 40% for $Al_2O_3$ and AlN barriers [5]. On the other hand, high RT TMR values in the range of 180% have been observed for Fe/MgO/Fe tunnel junctions, which is a larger TMR than the bulk Fe spin-polarization could provide [6].

In the pioneering work of Tedrow and Meservey, the spin-polarization of ferromagnets (FM) has been determined by measuring the tunneling from the FM to superconducting layers [7,8]. Other measurement techniques of the spin-polarization are restricted to bare (best vacuum) interfaces using tunneling point contacts and Andreev reflection [9], where the tip is again superconducting. Spin polarized photoemission has been used to determine the spin-polarization at the Fermi edge [10]. In order to obtain the spin-polarization related to transport phenomena, the photoemission energy resolution has to be in the order of 1 meV, which is very hard to perform [11]. Another way to measure the spin-polarization at buried interfaces is based on circular polarized electroluminescence [12,13], which requires an underlay light emitting diode structure. For example, a +30% spin-polarization has been measured with this technique for an Fe/$SiO_2$ tunnel barrier [14]. In all of the measurement types described above, very special sample conditions and preparation techniques are needed, which do not represent actual systems at ambient conditions used in spintronics investigations and/or applications. For the superconducting tunneling and the electroluminescence measurements low temperatures are mandatory. Additionally, technologically relevant interfaces - for example - in TMR devices are insulator/ferromagnet (FM) interfaces, which are buried. Therefore, all the above-mentioned surface sensitive methods could not be used. This prevents measurements on buried interfaces of real spintronic samples (devices) at ambient conditions (RT). By comparing all the different results, a large bandwidth of published spin-polarizations is visible. FM/Oxide interface spin-polarizations strongly depend on the type of used methods, sample preparation, and also temperatures. For more details about other possible techniques, we refer to Chapter 6 of Reference [8].

Therefore, comparing different experimental results is hard, due to interface-related details. For example, in a DFT calculation study of Fe tunneling tips, Fe (001) interface layers revealed negative spin-polarization [15], but bulk Fe polarizations are positive [9,16]. Also, point contacts revealed strong variations with respect to the bulk spin-polarization in the whole 3d series [15]. This shows the strong dependence of the spin-polarization with respect to the local coordination of the FM-TM atoms. Therefore, rough interfaces can show a variety of different spin-polarizations in real devices. As an example, for the half-metal $CrO_2$, where an almost 100% spin-polarization has been predicted [17] and also confirmed experimentally [18,19], a RT GMR effect for the $CrO_2$/$RuO_2$/$CrO_2$ system of up to 3000% has been predicted [20]. In contrast, only a GMR of about 0.2% has been observed [21], where the discrepancy has been explained by reversed and induced interface magnetizations at the $CrO_2/RuO_2$ interface [22,23].

Another observed phenomenon is based on the spin-polarization of simple pure Ni, where strong differences between calculated and experimental spin-polarizations have been observed. From band structure calculations, small s-band contributions are almost non-polarized, while the Ni 3d states are pure minority states [24]. Therefore, the spin-polarization for Ni should be negative and in the range of 25-80%, depending on coordination [15,24]. In contrast, reported Ni spin-polarizations are positive [7,9,25].

We present here a new and straight-forward way to directly measure the spin-polarization of buried interfaces by the local investigation of a voltage-dependent magneto-optical Kerr effect (VMOKE) signal where AC voltage has been applied at a doped $Si/SiO_2$/FM electrical capacitor-like device (see Fig. 1a). The general idea is quite simple. If we apply a voltage to this capacitor, we will either push electrons to the $SiO_2$/TM metal interface (negative voltage) or remove electrons from the interface (positive voltage). If the interface has no spin-polarization, the same amount of spin-up and spin-down electrons will be added (removed) to (from) the interface. Therefore, no change in the total sample magnetization will be observed. However, if the interface is majority spin polarized, the added electrons (negative voltage) will have the same spin-polarization and therefore increase the total sample magnetization, while the removal of electrons (positive voltage) will decrease the magnetization by reducing the number of electrons with parallel spin orientation. The opposite will happen with inverse interfacial spin-polarization, where added electrons will reduce the total magnetization. In the following, we will show that the VMOKE-effect could be quantitatively measured and compared to the number of voltage-related moved electrons, providing a quantitative determination of the spin-polarization, which is purely based on conduction electrons.

The magnetization is locally measured in longitudinal geometry (see Fig. 1) by a commercial focused laser beam MOKE system (Nanomoke 3: Durham Magneto Optics), where we additionally apply an AC voltage between the sample Au contact and the doped Si substrate. To fix the sign convention, positive voltage means positive at the Au contact and negative at the Si substrate. Utilizing a Lock-In amplifier to extract the AC-related Kerr response, which is then compared to the strength of the longitudinal MOKE hysteresis loop. In this manuscript, all AC-related amplitudes are given in root mean square (RMS) values. The experimental setup is sketched in Supplemental Material II, where Fig. S1b shows the optical pathway and the electrical connections of the Lock-In amplifier (Stanford Research SR 830 DSP) and the Nanomoke 3 system in longitudinal MOKE geometry, necessary to extract the AC (at 1399 Hz) VMOKE signal (voltage based MOKE detector difference), which will be normalized to the detector sum. The frequency of 1399 Hz has been almost arbitrarily chosen in a low-noise region. For our Nanomoke 3 system, we performed a detailed analysis of the voltage ratio of $(I_1-I_2)/(I_1+I_2)$ (see Supplemental Material II) with respect to the Kerr Rotation in degree. This voltage ratio to Kerr Rotation conversion factor is for our system 2.340 degree. For further details, see Supplemental Material II. High-purity gold (Au) and transition metals (Fe, Co, and Ni) targets were deposited by an in-house-designed Ar-beam sputtering device on a p-type Boron doped Si wafer with a nominal 50 nm thick $SiO_2$ layer on top. The thickness of the $SiO_2$ layer has been verified to be 52.4±1 nm by reflectometry, ellipsometry, and white light interferometry.

In Fig. 2, the field dependent VMOKE signal $\Delta m_V$, i.e. the AC voltage-dependent change in degree of Kerr rotation, and the related static field dependent Kerr rotation are presented. While the conventional MOKE signal simply represents the overall magnetization behavior, including saturation and coercive fields, the shown VMOKE signal is different. In the case of Fe and Co, the "polarity" is opposite with respect to the MOKE signal. This means a positive voltage is decreasing the MOKE signal. As already mentioned above, a positive voltage decreases the number of electrons at the interface. This decrease is related to a decrease in magnetization. Therefore, the Fe and Co spin-polarizations have to be majority spin polarized at the interface.

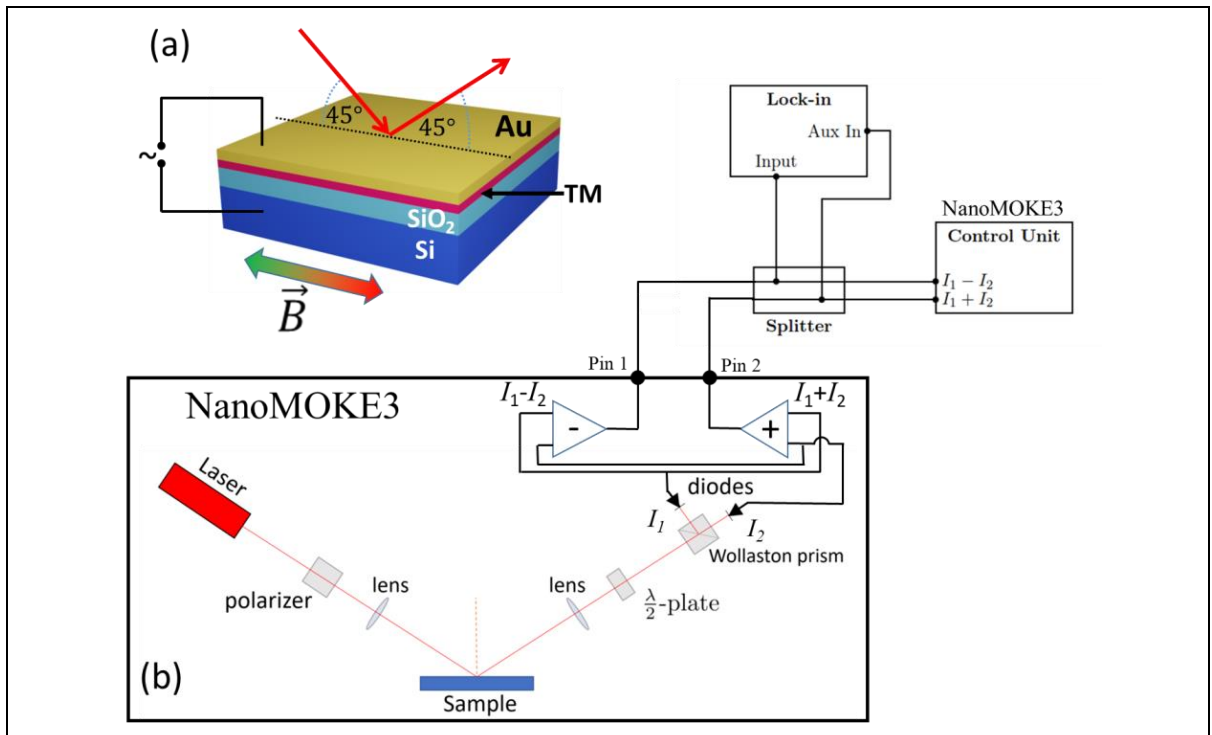

Fig. 1: a) Schematic of the electrical sample connections and the longitudinal MOKE scattering geometry with a 45° scattering angle. b) Sketch of the experimental setup, where the two frames labeled "NanoMOKE3" are rough schematics of the original MOKE from Durham Magneto Optics.

In the case of Ni, the VMOKE "polarity" is the same as for the MOKE signal, indicating that a positive voltage related reduction of interface electrons leads to an increase in magnetization, showing that the Ni spin-polarization is minority electron dominated. The general shape of the Ni VMOKE signal is also different compared to the hysteresis loop, which we will discuss later in more detail. Using the internal voltage conversion factors of the Nanomoke 3 (Supplemental Material II), the measured AC Lock-in voltage amplitude can be rescaled to provide an absolute value of the VMOKE signal, here in µ-degrees of Kerr rotation. The resulting order of the VMOKE magnitude is about four orders reduced compared to the conventional MOKE signal, and the so far obtained noise level is about 200 n degrees, which could be improved with increased measurement time. This demonstrates that we are able to observe VMOKE signal variations in the lower µ degree range.

In the following, a brief explanation will be given of how we determine absolute voltage-dependent changes of the sample magnetic moment (or magnetization) and also how to obtain absolute spin-polarizations at the interface. The general idea is as follows: First of all, the conventional MOKE signal is used as a reference for the total sample magnetic moment, also obtained by an additional conventional magnetometer. Here we used a Quantum Design MPMS 3 SQUID magnetometer. The corresponding SQUID-based magnetization curves are presented in Fig. 3. So, we multiply the VMOKE/MOKE signal ratio with the SQUID-determined saturation magnetic moment per area. This gives us an absolute voltage-dependent sample magnetic moment variation.

$$\Delta m_{Sample} \left[\frac{\text{emu}}{\text{Vm}^2}\right] = m_S^{SQUID} \left[\frac{\text{emu}}{\text{m}^2}\right] \cdot \frac{\Delta m_V[\text{mdeg}]/2}{M_S^{MOKE}[\text{mdeg}]} \cdot \frac{1}{U[\text{V}_{RMS}]}$$

Here $m_S^{SQUID}$ is the SQUID-determined sample saturation per area from zero to saturation, $M_S^{MOKE}$ is the MOKE related change in the polarization angle also from zero to saturation, $\Delta m_V$ is the AC voltage related change in the MOKE signal while changing from negative to positive saturation, and $U$ is the applied root mean square (RMS) AC voltage. The voltage-effect-based magnetic moment $\Delta m_V$ in units of degree Kerr rotation has to be normalized by 2, as all other quantities represent changes from zero to a positive maximum. Up to now, we have just determined the voltage-related change in saturation magnetic moment $\Delta m_{Sample}$ per voltage and area.

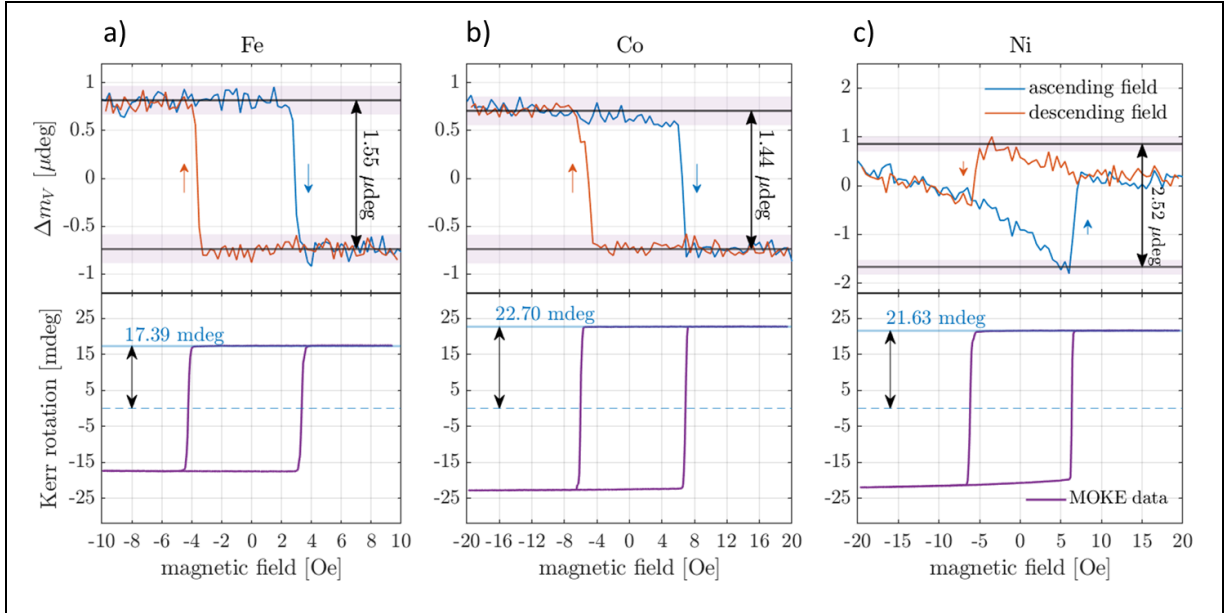

Fig. 2: The voltage-related changes $\Delta m_V$ as a function of the applied magnetic field are shown in the upper row, while the conventional Kerr effect rotations and their change from zero to saturation $M_S^{MOKE}$ are shown below. a), b), and c) represent the results for Fe, Co, and Ni, respectively.

Now we want to correlate this AC VMOKE signal with the number of electrons that are pushed to or away from the transition-metal/insulator interface. The general idea is as follows: If 100% spin-polarization is present at the Fermi level, or in other words, the interface has a half-metallic character, every electron will change the sample magnetic moment by 1µ$_B$ Bohr magneton. Here we assume an electron spin-only moment, neglecting orbital magnetic moments, which are usually much smaller compared to the spin moments, also at interfaces. The number of electrons could be calculated from the capacity of the junction or the displacement field, depending on the voltage, thickness, and dielectric constant at the junction. Here, the number of electrons moving to or away from the interface as a function of the voltage has been simply calculated from the displacement field. In Supplemental Material I further details for a 52.4 nm thick SiO$_2$ insulating layer are presented.

The total number of voltage-related electrons/area/volt times µ$_B$ would be the highest possible change $\Delta m_{100\%}$ in sample saturation, this is compared to the measured voltage change from above, which directly provides us with the spin-polarization at the interface:

$$P = \frac{-\Delta m_{Sample}}{\Delta m_{100\%}}$$

The minus sign is related to the fact that a positive voltage removes electrons from a majority spin (positive) polarized interface, which reduces the magnetization, while $\Delta m_{Sample}$ is negative. This ratio of the experimentally observed voltage-induced magnetic moment variation with respect to the calculated maximum voltage effect is directly related to the spin-polarization $P$ of the conducting electrons at the interface. $\Delta m_{100\%} = 3.72 \cdot 10^{-5} \frac{emu}{Vm^2}$ (see Supplemental Material I) is the same for all three samples as the substrate and the SiO$_2$ thickness are identical.

The related SQUID-determined saturation magnetic moments per area and the determined spin-polarizations for Fe, Co, and Ni are presented in Table 1.

Table 1: Fe, Co, and Ni SQUID-determined saturation magnetic moment per area, the VMOKE-based change in the magnetic moment per Volt and area, and the corresponding spin-polarizations.

| Element | $m_{sat}$ [EMU/m$^2$] | $\Delta m_{Sample}$ $\left[\frac{emu}{Vm^2}\right]$ | P [%] |
|---|---|---|---|
| Fe | 4.4±0.1 | $-2.2 \cdot 10^{-5}$ | 58±5 |
| Co | 3.8±0.1 | $-1.2 \cdot 10^{-5}$ | 32±3 |
| Ni | 0.9±0.1 | $5.8 \cdot 10^{-6}$ | -16±2 |

For a better understanding of the general concept, we have so far omitted an important experimental detail based on the superposition of a pure electro-optical (EO) effect, the so-called Pockels effect [26], to our magnetic VMOKE effect. This Pockels effect is a rotation of the electric field vector, similar to the magnetic Kerr effect, but not dependent on the magnetization and/or magnetic field and simply superimposed on our magnetic signal. Nevertheless, as the Pockels effect does not change with the magnetic field or sample magnetization, it is in principle just a constant offset. In Supplemental Material III we explain in detail how this Pockels effect manifests itself in our voltage dependent signal and how it is removed. In addition, we also demonstrate that the observed VMOKE effect is linear as a function of applied AC voltage (see Supplemental Material VI, Fig. S6).

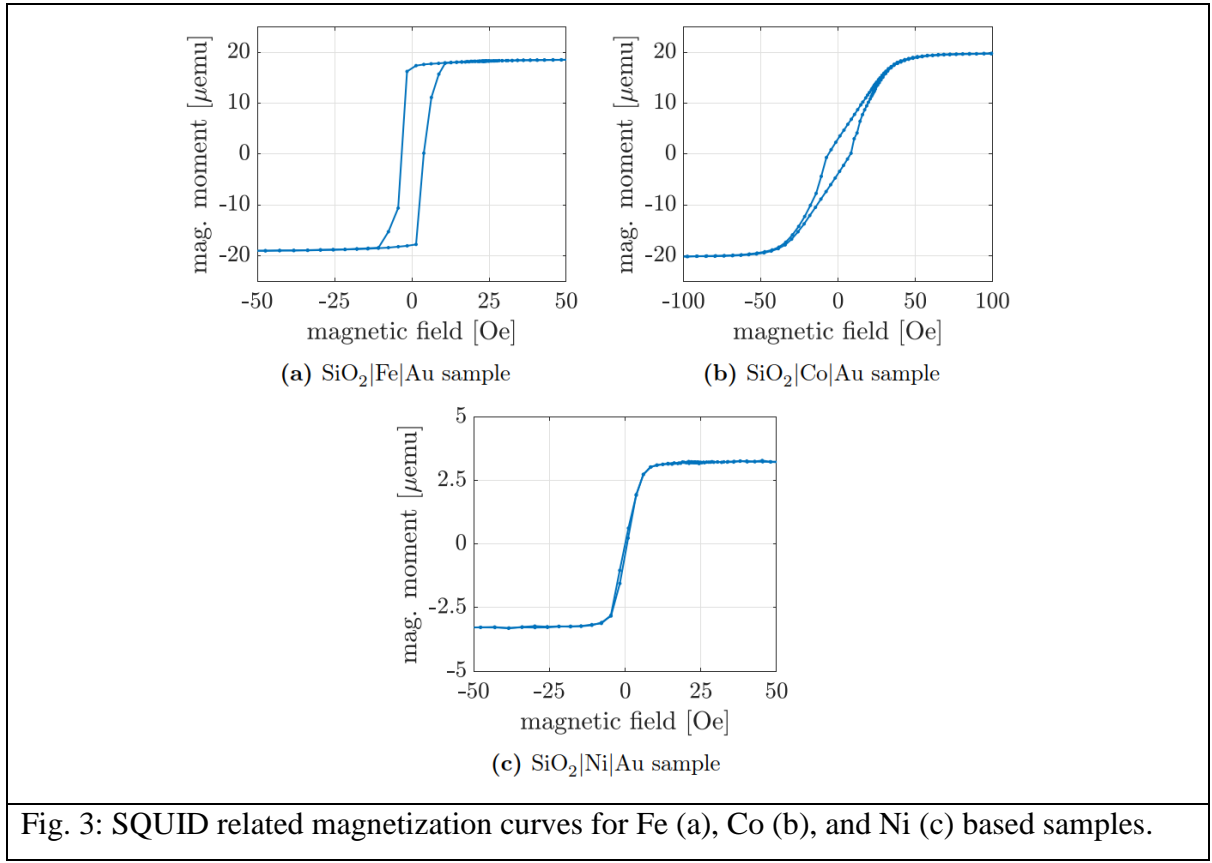

Fig. 3: SQUID related magnetization curves for Fe (a), Co (b), and Ni (c) based samples.

To exclude spin-orbit-coupling related effects, the current densities present here are, by many orders of magnitude, too small [27,28]. The insulating $SiO_2$ layer provides a total perpendicular junction resistance in the upper GOhm range (see Supplemental Material V), resulting in possible ohmic leakage currents of max. 1 nA. On the other hand, the RMS charging current at the 1399 Hz excitation is about 80 µA, which does not result in any ohmic heating or possible related spin-Seebeck effects [29,30].

In the case of Fe and Co, the shape of the VMOKE hysteresis is almost the same as the normal MOKE hysteresis, despite the opposite direction. While the MOKE signal 'jumps up' for positive magnetic fields, the VMOKE signal 'jumps down'. This is based on the change in saturation magnetic moment for positive applied voltage, which removes electrons (adds holes) at the interface. If we have majority spin-polarization, as is expected for Fe, any removed majority electron will decrease the saturation. Just from a simple inspection of the spin polarized density of states spin up and spin down ratio, as for example shown in [31], a bulk Fe spin polarized DOS ratio of about 77%/23% could be roughly determined, resulting in a net spin-polarization of about $P_{Fe}$=+55% at 0 K. The measured spin-polarization of +58% is close to the bulk band structure related value and to the photoemission-based value from Bush et al. +54% [32], and larger compared to other published tunneling-based references, such as 42-45% from Soulen et al. [9], or 40% from Meservey et al. [8] . In the case of Co, previous experimental observations have shown a majority spin-polarization of 42% from Soulen et al. [9], 35% from Meservey et al. [8], or Bush et al. +21% [32]. The Co spin-polarization of 32% determined here is close to the average value of the mentioned references, demonstrating again the quantitative validity of our approach. We still want to mention that the Co band structure provides an almost inverse majority to minority spin polarized DOS ratio of about 20%/80% with a resulting spin pol $P_{Co}$=-60% [31]. This again demonstrates that interfaces can provide significant variations in spin-polarization compared to bulk expectations.

As Ni has an almost filled 3d shell, it is expected to have a minority (negative) spin-polarization present at the Fermi level. In this case, a positive voltage (e.g., removing electrons from the interface) reduces the number of minority (antiparallel) spins and therefore increases the saturation magnetization. In strong contrast, all Ni spin-polarizations found in the literature were positive. The photoemission-based value from Bush et al. is +15% [32], as 43-46.5% from Soulen et al. [9], or 11% or 23% from Meservey et al. [7,8]. The VMOKE signal shown in Fig. 2c has a clear, visible 'jump up', reflecting negative, i.e., minority spin-polarization. The shape of the Ni VMOKE signal is quite different from the conventional MOKE signal, and the VMOKE signal vanishes at saturation. Following our explanation in Supplemental Material IV, where we compared the VMOKE Ni signal with transversal MOKE, we could provide a consistent picture by assuming that the voltage-related Ni interface electrons are almost perpendicularly polarized to the Ni bulk magnetization. As Ni is in direct contact with $SiO_2$, the interface atoms are at least slightly oxidized. Therefore, the perpendicular interface spin polarization is consistent with the well-known perpendicular coupling at NiO/Ni interfaces [33]. This also results in a vanishing projected Ni spin-polarization along the B-field direction for a saturated sample, which is also consistent with recently published voltage-related x-ray reflectometry data for almost equal samples [34].

We would like to mention that in our case, the VMOKE signal originates from the deepest FM layer at the SiO$_2$ interface. Therefore, the MOKE signal is reduced due to the visible (red) light penetration depth in metals since the laser beam must pass the Au and the TM layers to arrive and be reflected at the insulator/TM interface. The Au layer just reduces the total intensity and therefore also the overall MOKE signal, while the TM layer is damping the VMOKE signal from the interface with respect to the average MOKE signal. Simple estimates using the Fresnel equation and complex index of refractions show that this results for our TM film thicknesses in minor reductions of the spin-polarization of about 10% with respect to the total measured value, which is about to be the present error range. Further details about this effect need more systematic studies and will be discussed in a future contribution. Here, we want to say that very thin FM layers are important. Otherwise, the signal from the interface is reduced. Nevertheless, this could be quantitatively corrected by optical calculations. On the other hand, systematic studies on interface properties with the same FM layer stack on top will allow precise investigations of relative variations. We also want to emphasize, that, for example by increased integration time at saturation conditions, the absolute error of the determined spin polarization could be reduced to about 3-5%. By comparing it with the data scatter present in the actual literature, the error shown here is clearly on the lower end.

In summary, we could measure voltage-related changes in the MOKE signal for Fe, Co, and Ni thin films. By comparing the VMOKE signal with conventional MOKE hysteresis and SQUID results, we were able to obtain reasonable values for the spin-polarization. In the case of Ni, we could determine for the first time the expected minority band related negative spin-polarization, together with an unexpected almost perpendicular oriented interfacial spin-polarization. This method provides an easy and new way to measure absolute spin-polarization, purely related to conducting electrons in real buried layers and interfaces, where no tunneling artifacts will appear. As MOKE is a widely used method, this tool could be used by many groups in the world working on magnetism and spintronics on real samples at ambient conditions.

**Acknowledgments:** The authors thankfully acknowledge Bernd Ludescher for the help in preparing the samples, and Felix Groß, and Frank Schulz for fruitful discussions. We also thank Marko Burghard for valuable comments and improvements. We want to thank the Max-Planck-Society for financial support.

## References

[1] M. Julliere, *Tunneling between Ferromagnetic Films*, Phys Lett A **54**, 225 (1975).
[2] J. C. Slonczewski, *Conductance and Exchange Coupling of Two Ferromagnets Separated by a Tunneling Barrier*, Phys Rev B **39**, 6995 (1989).
[3] S. S. P. Parkin et al., *Exchange-Biased Magnetic Tunnel Junctions and Application to Nonvolatile Magnetic Random Access Memory (Invited)*, J Appl Phys **85**, 5828 (1999).
[4] S. S. P. Parkin, C. Kaiser, A. Panchula, P. M. Rice, B. Hughes, M. Samant, and S.-H. Yang, *Giant Tunnelling Magnetoresistance at Room Temperature with MgO (100) Tunnel Barriers*, Nat Mater **3**, 862 (2004).
[5] C. Kaiser, S. van Dijken, S.-H. Yang, H. Yang, and S. S. P. Parkin, *Role of Tunneling Matrix Elements in Determining the Magnitude of the Tunneling Spin Polarization of 3d Transition Metal Ferromagnetic Alloys*, Phys Rev Lett **94**, 247203 (2005).
[6] S. Yuasa, T. Nagahama, A. Fukushima, Y. Suzuki, and K. Ando, *Giant Room-Temperature Magnetoresistance in Single-Crystal Fe/MgO/Fe Magnetic Tunnel Junctions*, Nat Mater **3**, 868 (2004).
[7] P. M. Tedrow and R. Meservey, *Spin Polarization of Electrons Tunneling from Films of Fe, Co, Ni, and Gd*, Phys Rev B **7**, 318 (1973).
[8] R. Meservey and P. M. Tedrow, PHYSICS REPORTS Spin-Polarized Electron Tunneling, 1994.
[9] R. J. Soulen et al., *Measuring the Spin Polarization of a Metal with a Superconducting Point Contact*, Science (1979) **282**, 85 (1998).
[10] G. Busch, M. Campagna, and H. C. Siegmann, Spin-Polarized Photoelectrons from Fe, Co, And, n.d.
[11] R. Feder, *Polarized Electrons in Surface Physics* (WORLD SCIENTIFIC, 1986).
[12] H. J. Zhu, M. Ramsteiner, H. Kostial, M. Wassermeier, H.-P. Schönherr, and K. H. Ploog, *Room-Temperature Spin Injection from Fe into GaAs*, Phys Rev Lett **87**, 016601 (2001).
[13] N. Nishizawa, K. Nishibayashi, and H. Munekata, *Pure Circular Polarization Electroluminescence at Room Temperature with Spin-Polarized Light-Emitting Diodes*, Proceedings of the National Academy of Sciences **114**, 1783 (2017).
[14] C. H. Li, G. Kioseoglou, O. M. J. van 't Erve, P. E. Thompson, and B. T. Jonker, *Electrical Spin Injection into Si(001) through a SiO2 Tunnel Barrier*, Appl Phys Lett **95**, 172102 (2009).
[15] P. Ferriani, C. Lazo, and S. Heinze, *Origin of the Spin Polarization of Magnetic Scanning Tunneling Microscopy Tips*, Phys Rev B **82**, 054411 (2010).
[16] M. Häfner, J. K. Viljas, D. Frustaglia, F. Pauly, M. Dreher, P. Nielaba, and J. C. Cuevas, *Theoretical Study of the Conductance of Ferromagnetic Atomic-Sized Contacts*, Phys Rev B **77**, 104409 (2008).
[17] M. A. Korotin, V. I. Anisimov, D. I. Khomskii, and G. A. Sawatzky, *CrO2: A Self-Doped Double Exchange Ferromagnet*, Phys Rev Lett **80**, 4305 (1998).
[18] J. S. Parker, S. M. Watts, P. G. Ivanov, and P. Xiong, *Spin Polarization of CrO2 at and across an Artificial Barrier*, Phys Rev Lett **88**, 196601 (2002).
[19] A. Anguelouch, A. Gupta, G. Xiao, D. W. Abraham, Y. Ji, S. Ingvarsson, and C. L. Chien, *Near-Complete Spin Polarization in Atomically-Smooth Chromium-Dioxide Epitaxial Films Prepared Using a CVD Liquid Precursor*, Phys Rev B **64**, 180408 (2001).
[20] A. M. Bratkovsky, *Tunneling of Electrons in Conventional and Half-Metallic Systems: Towards Very Large Magnetoresistance*, Phys Rev B **56**, 2344 (1997).
[21] G. X. Miao, A. Gupta, H. Sims, W. H. Butler, S. Ghosh, and G. Xiao, *Giant Magnetoresistive Structures Based on CrO2 with Epitaxial RuO2 as the Spacer Layer*, J Appl Phys **97**, 10C924 (2005).
[22] K. Zafar, P. Audehm, G. Schütz, E. Goering, M. Pathak, K. B. Chetry, P. R. LeClair, and A. Gupta, *Cr Magnetization Reversal at the CrO2/RuO2 Interface: Origin of the Reduced GMR Effect*, Phys Rev B **84**, 134412 (2011).
[23] K. B. Chetry, H. Sims, W. H. Butler, and A. Gupta, *Electronic and Magnetic Structure of CrO2/RuO2 Interfaces*, Phys Rev B **84**, 054438 (2011).
[24] H. Eckardt and L. Fritsche, *The Effect of Electron Correlations and Finite Temperatures on the Electronic Structure of Ferromagnetic Nickel*, Journal of Physics F: Metal Physics **17**, 925 (1987).
[25] R. Meservey and P. M. Tedrow, *Spin-Polarized Electron Tunneling*, Phys Rep **238**, 173 (1994).

[26] F. Pockels, *Ueber Den Einfluss Elastischer Deformationen, Speciell Einseitigen Druckes, Auf Das Optische Verhalten Krystallinischer Körper, Dissertation, Göttingen, 1889.* (VDM-Verlag Dr. Müller, Saarbrücken, 2008).
[27] J. Sinova, D. Culcer, Q. Niu, N. A. Sinitsyn, T. Jungwirth, and A. H. MacDonald, *Universal Intrinsic Spin Hall Effect*, Phys Rev Lett **92**, (2004).
[28] L. Liu, R. A. Buhrman, and D. C. Ralph, Review and Analysis of Measurements of the Spin Hall Effect in Platinum, n.d.
[29] G. E. W. Bauer, E. Saitoh, and B. J. Van Wees, *Spin Caloritronics*, Nature Materials.
[30] K. Uchida, S. Takahashi, K. Harii, J. Ieda, W. Koshibae, K. Ando, S. Maekawa, and E. Saitoh, *Observation of the Spin Seebeck Effect*, Nature **455**, 778 (2008).
[31] J. M. D. Coey, *Materials for Spin Electronics*, in (2001), pp. 277–297.
[32] G. Busch, M. Campagna, and H. C. Siegmann, *Spin-Polarized Photoelectrons from Fe, Co, and Ni*, Phys Rev B **4**, 746 (1971).
[33] J. Nogués and I. K. Schuller, *Exchange Bias*, J Magn Magn Mater **192**, 203 (1999).
[34] S. E. Ilse, G. Schütz, and E. Goering, *Voltage X-Ray Reflectometry: A Method to Study Electric-Field-Induced Changes in Interfacial Electronic Structures*, Phys Rev Lett **131**, 036201 (2023).

## Materials

High-purity Gold (Au) and transition metals (Fe, Co, and Ni) targets were deposited by an in-house-designed Ar-beam sputtering device on a p-type Boron doped Si wafer, with a nominal 50 nm thick SiO2 layer on top. The thickness of the SiO2 layer has been verified to be 52.4±1 nm by reflectometry, ellipsometry, and white light interferometry.

## Methods

MOKE: We used a commercial Nanomoke 3 Moke System from Durham Magneto Optics (DMO) in Combination with a Stanford Research SR 830 DSP lock in amplifier. We used the standard quadrupolar DMO electromagnet, where only the longitudinal field component has been excited. The maximum achievable magnetic field is 0.12 T. For more details, see section Supplemental Material II.

Magnetometry: For magnetometry, we used a commercial MPMS3 VSM SQUID system from Quantum Design.

# Supplemental Material to
# "Direct determination of the absolute spin-polarization at ferromagnet/insulator interfaces using voltage dependent MOKE signals."

**I: Displacement field and maximum voltage related change in the magnetization**

The insulating layer is made of commercially thermally oxidized silicon with a given thickness of $d_{SiO2}$ = 52.4 nm. On top of the ferromagnetic layer, a 5 nm thick Au layer is deposited to prevent the metal from oxidizing, while providing decent electrical contact. The system describes a capacitor like structure (Figure 1 in the main text), and in order to calculate the number of electrons shifted towards the interface of the insulating layer, the electrical displacement field has to be determined. Assuming the applied voltage is $U$ = 10 V, the electric field is:

$$E = \frac{10\text{V}}{52.4\text{nm}} = 1.91 \cdot 10^8 \frac{\text{V}}{\text{m}}$$

Using the dielectric constant of $SiO_2$ of $\varepsilon_r = 3.8$ [1] the displacement field D in units of Coulomb/area or number of electrons/area as

$$D = \varepsilon_0 \varepsilon_r E = 6.42 \cdot 10^{-3} \frac{\text{C}}{\text{m}^2} = 4.01 \cdot 10^{16} \frac{\text{e}^-}{\text{m}^2}$$

If we divide the displacement again by the applied voltage $U$ = 10 V we obtain the number of electrons per area and voltage moving to or from the interface, depending on polarity. For our 52.4 nm-thick samples, we get

$$4.01 \cdot 10^{15} \frac{\text{e}^-}{\text{m}^2 V}$$

We also measured the electronic capacity for all our samples and obtained very similar results, but with larger errors, due to area determination errors and parasitic capacities, where the latter is hard to subtract as these parasitic capacities are in the same order or even larger compared to the sample capacities (not shown). Therefore, we decided to stick with the theoretical calculation. For a 100% spin-polarized interface, all of these electrons contribute, with the same spin of ½ and g-factor of 2, to a change in sample magnetization. Therefore, every electron changes the magnetization by 1 $\mu_B$ (Bohr magneton). A single Bohr magneton has $9.274 \cdot 10^{-21}$ EMU we can easily calculate the 100% spin-polarization change $\Delta m_{100\%}$ in sample magnetization per area and voltage as

$$\Delta m_{100\%} = 4.01 \cdot 10^{15} \frac{\text{e}^-}{\text{m}^2 V} \cdot 9.274 \cdot 10^{-21} \frac{\text{EMU}}{\text{e}^-} = 3.72 \cdot 10^{-5} \frac{\text{EMU}}{\text{m}^2 V}$$

As described in the main text, we measured the change in voltage-related saturation magnetic moment with the MOKE system. Therefore, the spin-polarization could be easily calculated by the ratio between the observed change in magnetization and the calculated 100% polarization-related change in magnetization, as

$$P = \frac{-\Delta m_{Sample}}{\Delta m_{100\%}}$$

The negative sign is as follows: First of all, we assume a less than half-filled 3d shell, where only majority electrons are occupied, and the lowest unoccupied electron states are also pure majority-like. If we now apply a positive voltage to this 3d layer, we introduce holes at the interface by simply removing electrons. These electron spins are missing, and the magnetization is reduced. Measuring this effect as a function of field provides us with a hysteresis loop with a jump to smaller magnetization at positive magnetic fields (see Fig. 2 in the main text). The sign of the voltage-related hysteresis 'jump' is negative. If we define positive polarization as being parallel to the majority electrons, we introduce a negative sign in the above given formula for the polarization.

**II: Setup and Intensity vs. Kerr rotation relation**

The general setup is shown in main Fig. 1 and in Fig. S1a, where Fig.S1b shows the electrical connections of the Lock-In amplifier (Stanford Research SR 830 DSP) and the Nanomoke 3 system, necessary to extract the AC VMOKE signal (Kerr detector difference), which is normalized to the detector sum. The Nanomoke 3 system calculates inside its own control unit (see Fig.S1b) the Kerr rotation from the asymmetry ratio between a two-channel difference ($I_1-I_2$) normalized by the reflectivity signal ($I_1+I_2$). Where $I_1$ and $I_2$ are voltages proportional to the light intensities measured behind the Wollaston prism (see Fig. 1b). To our knowledge, the sum signal is generated by an additional internal photodetector for optimal noise reduction. Nevertheless, both signals are transmitted from the Nanomoke 3 to its control unit by simple voltages at pins 1 and 2. We measure the AC voltage-related changes of the difference signal by our lock in amplifier (Stanford Research SR 830 DSP), as shown in Fig. S1a. In addition, the sum signal is also monitored by the auxiliary input of the lock in amplifier.

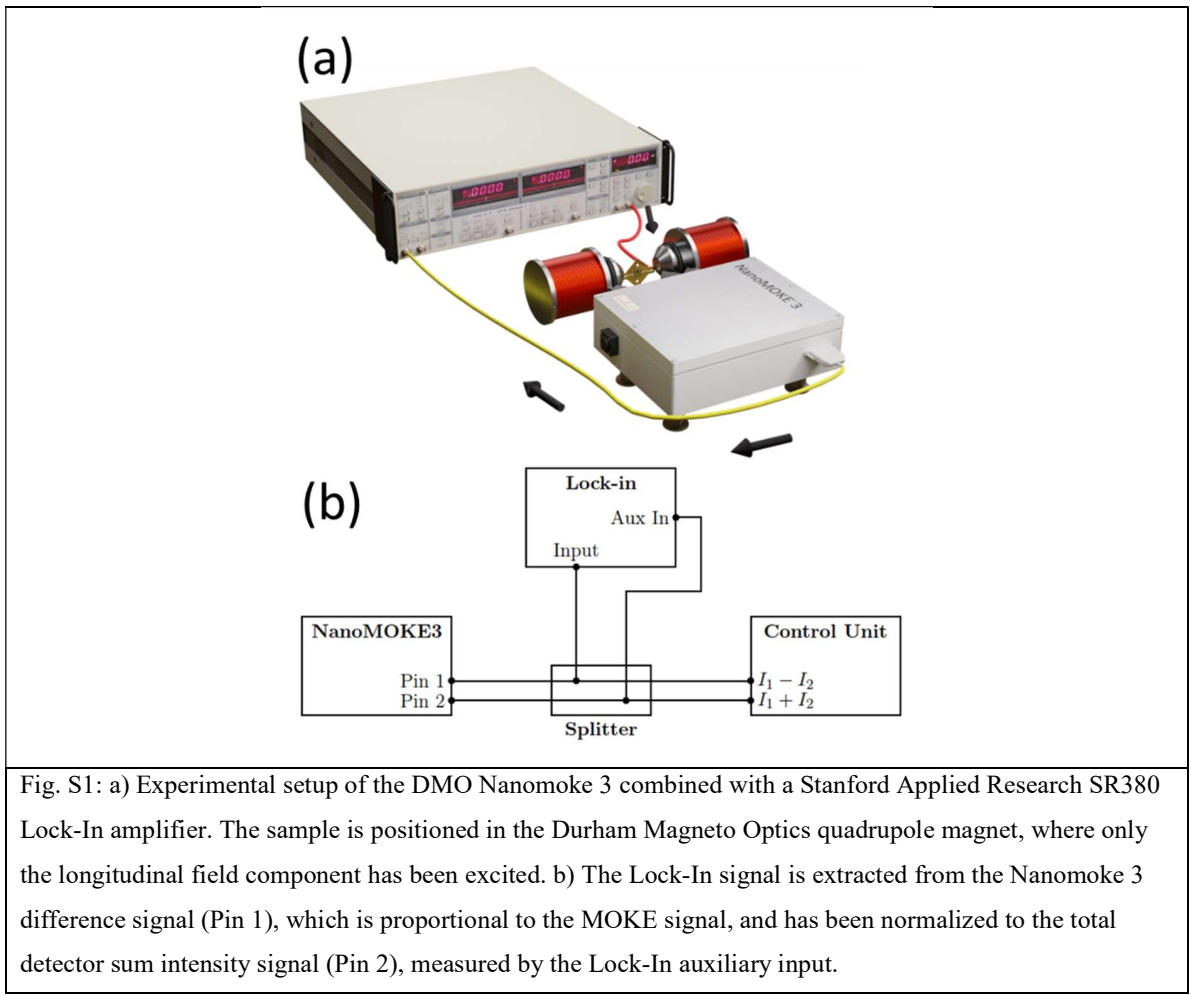

Fig. S1: a) Experimental setup of the DMO Nanomoke 3 combined with a Stanford Applied Research SR380 Lock-In amplifier. The sample is positioned in the Durham Magneto Optics quadrupole magnet, where only the longitudinal field component has been excited. b) The Lock-In signal is extracted from the Nanomoke 3 difference signal (Pin 1), which is proportional to the MOKE signal, and has been normalized to the total detector sum intensity signal (Pin 2), measured by the Lock-In auxiliary input.

So the Kerr angle $\theta$ is calculated in the Nanomoke control unit from the channel asymmetry by a simple proportionality constant $c$ as follows:

$$\theta = c \cdot \frac{I_1 - I_2}{I_1 + I_2} \; .$$

In order to determine this proportionality constant, we could vary the signal offset in the Nanomoke system using the half-wave plate, resulting in an artificial Kerr signal. This Kerr rotation given by the Nanomoke software (LxPro3) is depicted in Fig. S2 as a function of the above-mentioned voltage ratio taken from the NanoMOKE3. Utilizing a simple linear fit, we determined the conversion factor slope as $c$=2.340 degrees between the voltage asymmetry and the Kerr angle $\theta$. The unit is just degrees, as the voltage ratio is dimensionless.

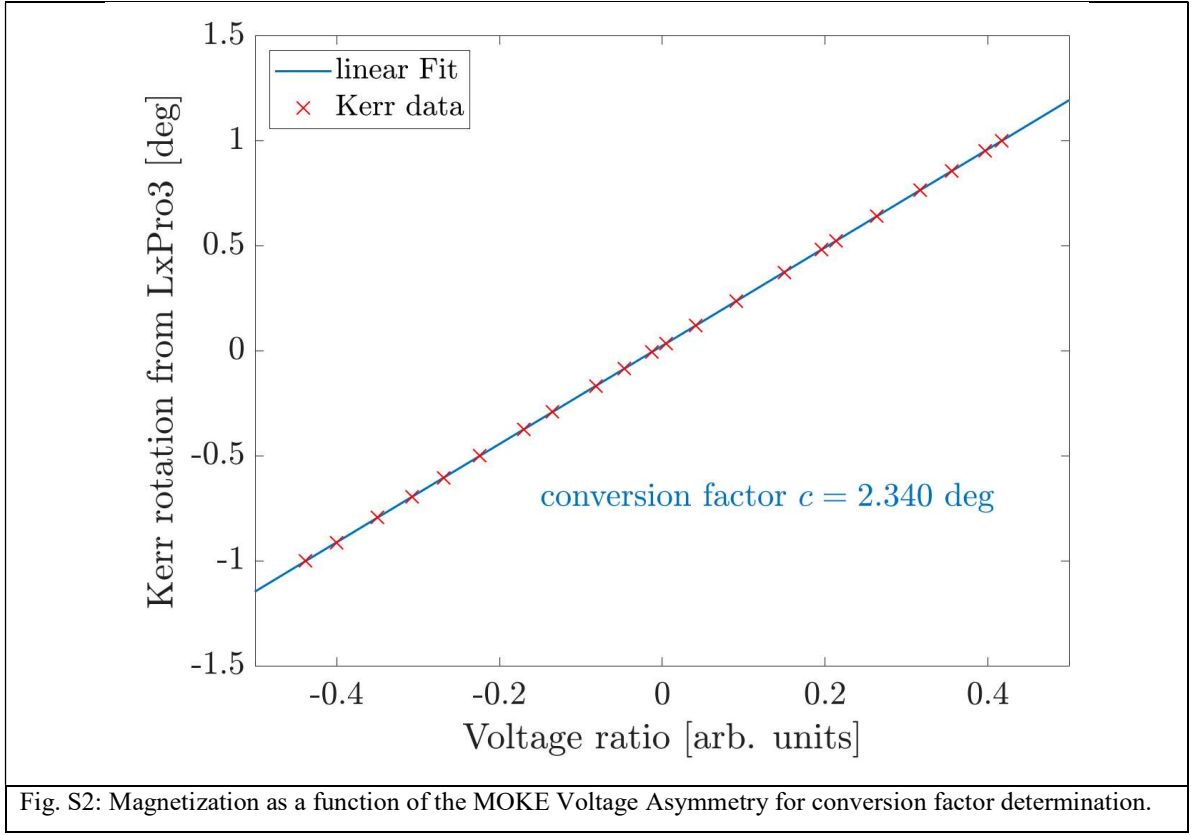

Fig. S2: Magnetization as a function of the MOKE Voltage Asymmetry for conversion factor determination.

In order to obtain our voltage-related Kerr signal, also in units of degree, we simply measure the in-phase component corresponding to our AC sample voltage using the Lock-In voltage $V_{LockIn}$, as shown in Fig. S1, normalized by the intensity sum-related voltage and multiplied by the determined conversion factor:

$$m_V = 2.340 \text{ [deg]} \cdot \frac{V_{LockIn}}{(I_1 + I_2)}$$

Again $I_1 + I_2$ is the intensity sum given in Volt!

**III: Details in the measurement sequence and removal of the superimposed Pockels effect**

In addition to our VMOKE effect, the electro-optical (EO) or Pockels effect [2] is superimposed on our measured Lock-In signal. This can result in strong rotations of the light polarization in thin film systems [3,4]. This is a non-magnetism-related voltage effect, that is based on voltage-dependent interface symmetry breaking as the electrical field has a well-defined direction and sign and results in an up or down shift of the VMOKE lock in signal as a function of applied voltage as indicated in Fig. S3. In the following, we explain how this Pockels effect will modify our VMOKE signal and how we removed this effect from our VMOKE signal.

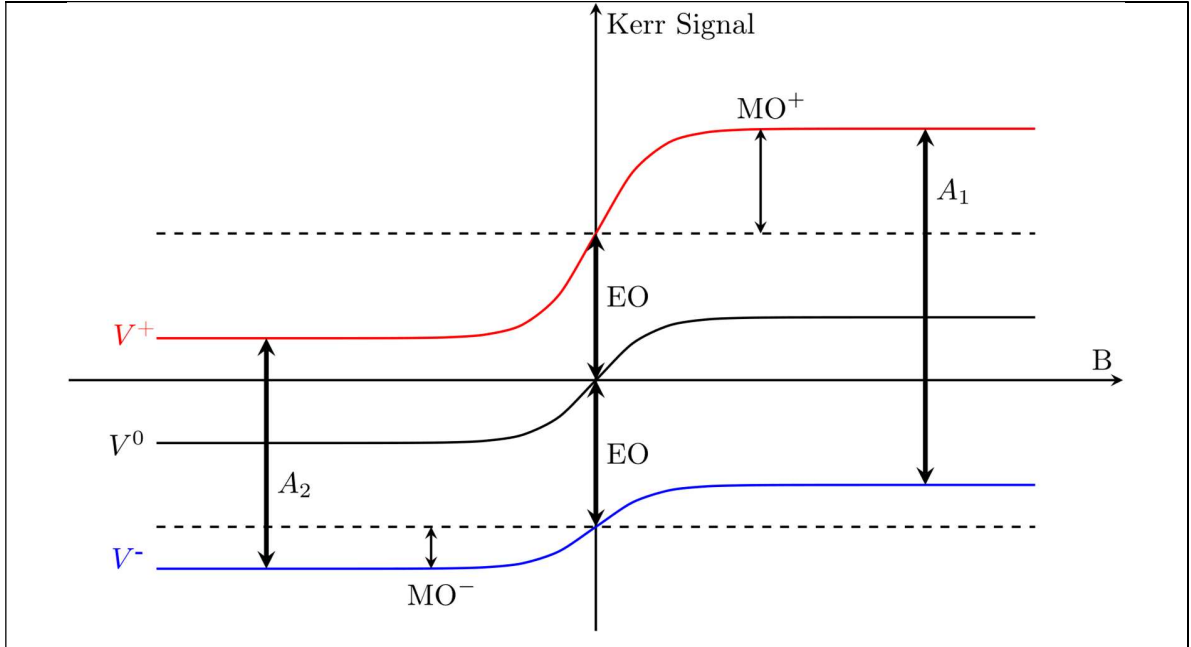

Fig. S3: Shown is a schematic of the effects present when applying a voltage while measuring the hysteresis via the MOKE effect. The Kerr signal is displayed as a function of the applied magnetic field. The abbreviation EO stands here for the electro-optical effect, which leads to an offset shift of the hysteresis along the Kerr axis. MO± stands for the magneto-optical change in saturation magnetization depending on the voltage polarity.

In Fig. S3, we show three different idealized MOKE curves for the condition with no applied voltage ($V^0$) and with applied positive and negative voltages ($V^+$ and $V^-$). In reality, the difference between $V^+$ and $V^-$ is exactly our magnetic field-dependent lock-in amplitude. For better visibility, the magnetic effect is amplified by orders of magnitude. In this example, the positive voltage (red curve) shifts the hysteresis curve to higher positive Kerr angles, while the negative voltage (blue curve) behaves in the opposite way. In our example, the positive voltage increases the MOKE effect (i.e. increases the sample magnetization), while the negative voltage decreases the MOKE effect. This corresponds to a minority spin-polarization, where a decrease in electrons at the Fermi level increases the saturation magnetization (see main text). The amplitude difference for positive and negative voltage is defined as $A_1$ in a positive magnetic field configuration and $A_2$ in negative field direction. The lock-in, measuring sinusoidal AC voltage, directly measures the amplitudes $A_1$ and $A_2$ between the hysteresis with positive applied voltage and the hysteresis with negative applied voltage. Because the electro-optical effect isn't influenced by the magnetic field, its amplitude is constant in each direction of the magnetic field. From Fig. S3, the following two equations are derived for the measured amplitude:

$$A_1 = 2 \cdot EO - MO^- + MO^+$$
$$A_2 = 2 \cdot EO + MO^- - MO^+$$

Here, $MO^\pm$ relates to the magnetic effect, respectively, of the positive and negative applied voltages. If the sample is in magnetic saturation, the magnetic effect is solely dependent on the applied voltage.

While subtracting the EO effect, we obtain on the positive (negative) B field side the following VMOKE amplitudes: VMOKE$^+$ and VMOKE$^-$, respectively:

$$VMOKE^+ = A_1 - 2 \cdot EO = MO^+ - MO^-$$
$$VMOKE^- = A_1 - 2 \cdot EO = -(MO^+ - MO^-)$$

As we measure at each magnetic field the full voltage-dependent effect, our Lock-In amplitude contains the VMOKE effect ($MO^+ - MO^-$) and $2 \cdot EO$ at every single magnetic field-dependent data point (calculated here for the saturation amplitudes $A_1$ and $A_2$). Or, in other words, our VMOKE Lock-In signal is just shifted to positive values by a magnetic field-independent value of $2 \cdot EO$. Following the determined equations used to remove the Pockels effect from our measurement, our measurement sequence is as follows: We apply a magnetic field, as we would in conventional hysteresis. We measure the corresponding lock-in signal for every applied field. After a half hysteresis loop (from -B to +B), the average lock-in value is calculated and subtracted from every lock-in data point in each loop to remove the EO effect. This has been performed for up and down loops separately. As the VMOKE signal is quite small, we have to repeat the measurement until we obtain a sufficient signal/noise (S/N) ratio, as presented in the main text. In our case, each loop consists of 400 B-field-related data points, and every loop has been repeated 50 times to get sufficient S/N. Every single VMOKE hysteresis loop takes about 2 days in total.

**IV: Ni VMOKE signal compared to transversal MOKE**

Due to the unusual shape of the Ni VMOKE signal, we also investigated the transversal MOKE signal of the Ni sample, i.e., the reflected intensity of the laser beam as a function of the magnetic field, which reflects the transversal component of the total sample magnetization, as shown in Fig. S4a. This shows great similarity to the VMOKE measurement, which indeed infers perpendicular interfacial in-plane magnetization. The shape of this transversal Kerr signal could be interpreted as a superposition of a signal corresponding to the Ni VMOKE signal and a conventional hysteresis. As we have a clear jump from positive to negative saturation and vice versa, a simple square-shaped hysteresis would be expected, which is not going to zero due to a small difference between the direction of the applied field and the MOKE detection direction and/or slight tilted easy axis orientations with respect to the field. And of course, this signal is not the same as the Ni VMOKE response, as here the total transverse total sample signal is shown, as the VMOKE signal is just the interface, and only a fraction of the interface signal is related to moving electrons. This suggests that the spin-polarization at the Ni interface is in saturation almost perpendicular to the sample magnetization, providing almost vanishing effects in saturation and maximum strength close to the switching field, where the sample magnetization also tilts away from the parallel configuration. Therefore, we can identify an almost perpendicularly magnetized interface magnetization with respect to the sample magnetization.

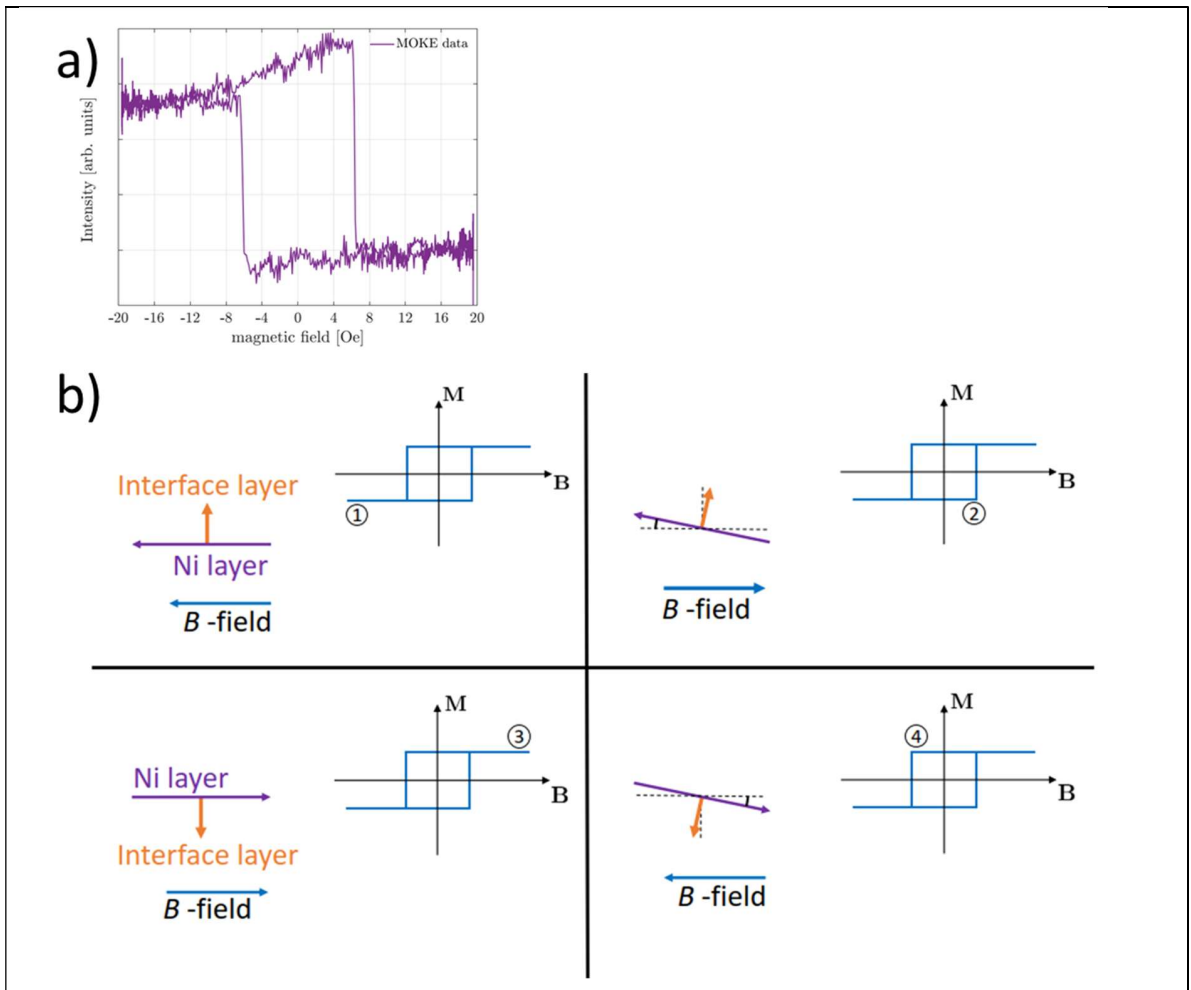

Fig. S4: a) Transverse MOKE signal of the Ni sample. b) The switching behavior of two mutually perpendicularly magnetized layers (here the in-plane Ni layer and the in-plane interface layer) under an applied magnetic field is shown in a schematic representation. The field sweep is ordered with [index (1)-(4)]. In positive and negative saturation, (1) and (3), the interface layer is completely perpendicularly magnetized in contrast to the Ni layer. At the switching points of the hysteresis (2) and (4), the applied field forces the interface layer and the Ni layer to tilt in the field direction, therefore creating a parallel net magnetic moment of the interface layer, which can be seen in the hysteresis of a longitudinal MOKE measurement.

One possible explanation for this perpendicular interface is based on the formation of an NiO like layer at the interface. For NiO, which is an antiferromagnet, it has been shown that the interface magnetization is often oriented perpendicular with respect to the metallic Ni ferromagnet. In itself, NiO is an antiferromagnet (AFM) and a coupling between AFM|FM (here the Ni layer) could lead to exchange bias, which occurs in thin films also in perpendicular coupling [5] and, according to [6], can show relatively strong exchange coupling. This is not present here since the NiO layer here is not a real AFM. From our results, the layer shows a non-AFM-like uncompensated net magnetization in perpendicular direction, otherwise, we would not see a VMOKE effect. This is also consistent with the absence of exchange bias. On the other hand, for exchange bias, a field cooling procedure from elevated temperatures is necessary, which has been omitted in order to not change the interfaces due to hot sample-induced atomic intermixing phenomena. Furthermore, it can be seen from Fig. S4a that when the magnetic field is fully passed, the spins do not flip back over the same path, but over a full 360 rotation, as illustrated in Fig. S4b.

## V. Electrical characterization

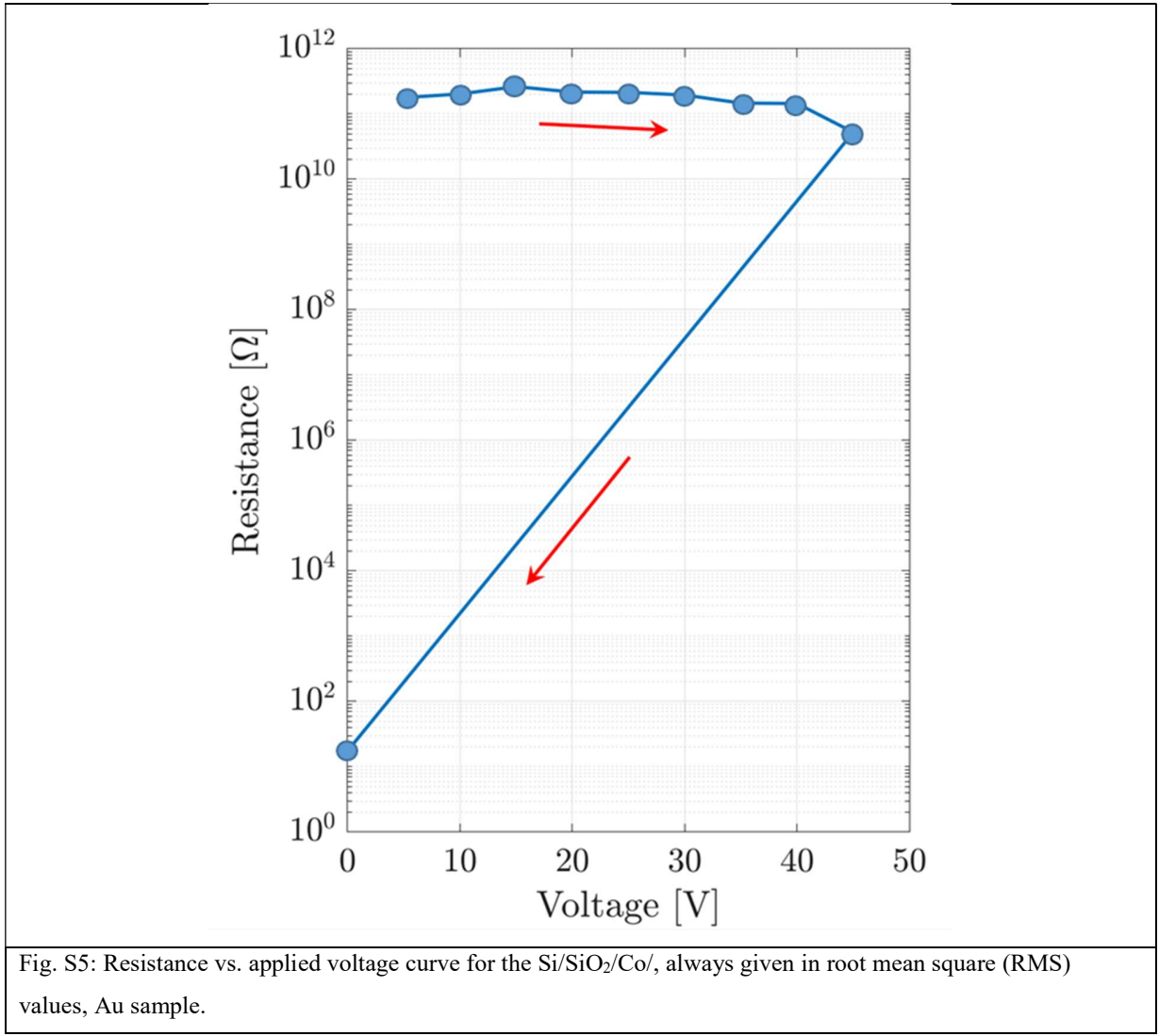

Fig. S5: Resistance vs. applied voltage curve for the Si/SiO$_2$/Co/, always given in root mean square (RMS) values, Au sample.

In Fig. S5, a typical sample resistance is shown as a function of the applied voltage. This behavior is almost independent of the TM used as a layer on top of the SiO$_2$. A typical breakdown behavior is visible. For voltages up to a threshold value of about 45 V, the resistance is above 10 GOhm. Then a breakthrough appears, with a corresponding drop in resistance down to values of less than 100 Ohm. In our VMOKE measurements, we use an RMS voltage of about 10 V, which corresponds to a peak voltage of about 14 V. This results in electric currents of less than 0.14 nA and dissipated peak power of less than 2 nW.

## VI: VMOKE linearity check

As this is a new method, we also want to demonstrate the linearity of the VMOKE effect as a function of the applied RMS voltage, which is shown in Fig. S6. Here, we just measured the VMOKE signal only at two magnetic fields for positive and negative saturation and plotted the direct difference $\Delta m_V$ as a function of the applied RMS voltage $V_{RMS}$. Within the error bars, a clear proportionality could be observed.

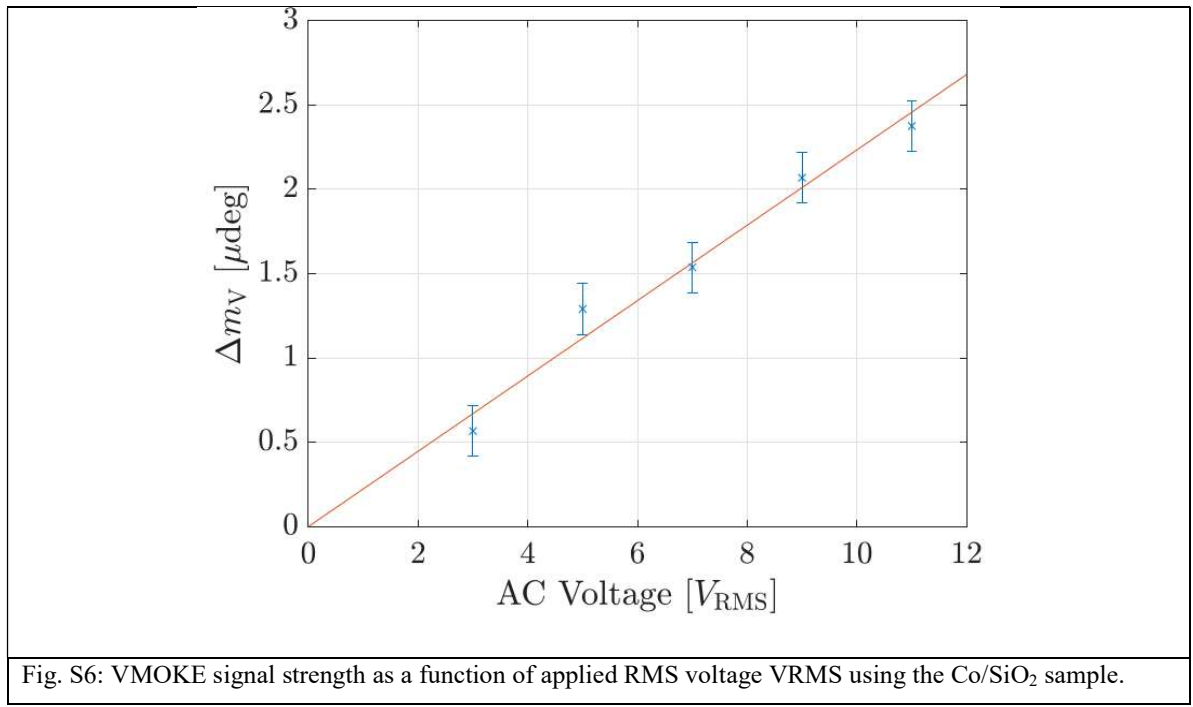

Fig. S6: VMOKE signal strength as a function of applied RMS voltage VRMS using the Co/SiO$_2$ sample.

**Supplemental Material References:**

[1]  A. R. von Hippel and S. O. Morgan, *Dielectric Materials and Applications*, J Electrochem Soc **102**, 68C (1955).
[2]  F. Pockels, *Ueber Den Einfluss Elastischer Deformationen, Speciell Einseitigen Druckes, Auf Das Optische Verhalten Krystallinischer Körper, Dissertation, Göttingen, 1889.* (VDM-Verlag Dr. Müller, Saarbrücken , 2008).
[3]  P. Damas, X. Le Roux, D. Le Bourdais, E. Cassan, D. Marris-Morini, N. Izard, T. Maroutian, P. Lecoeur, and L. Vivien, *Wavelength Dependence of Pockels Effect in Strained Silicon Waveguides*, Opt Express **22**, (2014).
[4]  D. Hayama, K. Seto, K. Yamashita, S. Yukita, T. Kobayashi, and E. Tokunaga, *Giant Pockels Effect in an Electrode-Water Interface for a "Liquid" Light Modulator*, OSA Contin **2**, 3358 (2019).
[5]  J. Nogués and I. K. Schuller, *Exchange Bias*, J Magn Magn Mater **192**, 203 (1999).
[6]  N. C. Koon, *Calculations of Exchange Bias in Thin Films with Ferromagnetic/Antiferromagnetic Interfaces*, Phys Rev Lett **78**, 4865 (1997).